\documentclass[twocolumn]{svjour3}          
\smartqed  
\usepackage{graphicx}
\begin{document}

\title{Tuning Crystal Field Potential by Orbital Dilution \\
in Strongly Correlated $d^4$ Oxides
\thanks{This work
is supported by the Foundation for Polish Science through
the IRA Programme co-financed by EU within \mbox{SG OP Programme.}}
}

\titlerunning{Tuning Crystal Field Potential by Orbital Dilution in $d^4$ Oxides}

\author{Wojciech Brzezicki \and Filomena Forte \and Canio Noce \and Mario Cuoco \and Andrzej M. Ole\'s
}


\institute{Wojciech Brzezicki  \at
International Research Centre MagTop, Institute of Physics PAS, Aleja Lotnik\'ow 32/46, PL-02668 Warsaw, Poland
\and
Filomena Forte \and Canio Noce \and Mario Cuoco \at
             CNR-SPIN, IT-84084 Fisciano (SA), Italy;\\
             Dipartimento di Fisica \textquotedblleft{}E. R. Caianiello\textquotedblright{},\\
             Universit\'a di Salerno, IT-84084 Fisciano (SA), Italy
           \and
           Andrzej M. Ole\'s \at
             a.m.oles@fkf.mpg.de \\
             Max Planck Institute for Solid State Research,\\
             Heisenbergstrasse 1, D-70569 Stuttgart, Germany;\\
             Marian Smoluchowski Institute of Physics, Jagiellonian Univ.
             Prof. S. \L{}ojasiewicza 11, PL-30348 Krak\'ow, Poland
}

\date{Received: 23 June 2019 / Accepted: 7 November 2019}

\maketitle

\begin{abstract}
We investigate the interplay between Coulomb driven orbital order
and octahedral distortions in strongly correlated Mott
insulators due to orbital dilution, i.e., doping by metal ions
without an orbital degree of freedom. In particular, we focus on
layered transition metal oxides and study the effective spin-orbital
exchange due to $d^3$ substitution at $d^4$ sites. The structure of
the $d^3-d^4$ spin-orbital coupling between the impurity and the
host in the presence of octahedral rotations favors a distinct type
of orbital polarization pointing towards the impurity and outside
the impurity--host plane. This yields an effective lattice potential
that generally competes with that associated with flat octahedra and,
in turn, can drive an inversion of the crystal field interaction.

\end{abstract}

\section{Introduction}

Transition metal oxides (TMOs) are fascinating materials where several
quantum degrees of freedom (i.e., spin, orbital, charge, lattice) are
intertwined, and require to be treated on equal footing both from a
fundamental point of view as well as for developing and enhancing
applications in the areas of oxide electronics~\cite{Col19}.
The competition of different types of ordered states is ubiquitous in
strongly correlated TMOs, mainly arising from the complex nature of the
spin-charge-orbital couplings where frustrated Coulomb driven exchange
competes with the kinetic energy of charge carriers.

A crucial step for accessing the emergent phenomena of correlated TMOs
with spin-orbital-charge coupled degrees of freedom is the understanding
of the undoped regime \cite{Kho14}, where the low-energy physics and
spin-orbital order are dictated by effective spin-orbital superexchange \cite{Tok00,Kug82,Fei97,Fei99,Kha05,Ole05,Nor08,Nor11,Cha11,Karlo,Ole12,Brz19}.
In undoped $3d$ Mott insulators, for instance, large local Coulomb
interactions localize electrons and the coupling between transition
metal ions is controlled by a low-energy spin-orbital superexchange
introduced first by Kugel and Khomskii \cite{Kug82}. When multi-orbital
degrees of freedom are included, the enhanced quantum fluctuations for
$S=1/2$ compounds can result in destroying the long range order
\cite{Fei97}. On the other hand, spin-orbital entanglement in
superexchange models may lead to exotic novel types of magnetic order
\cite{Brz12}, or stabilize topological order in the ground state and
in excited states \cite{Brz14}.

Long range order in both spin and orbital sector develops in perovskite
lattices when spin fluctuations are weaker for spin $S$ larger than $1/2$.
The fate of spin-orbital order is, however, strongly tied to the
character of the orbital degrees of freedom that emerges when electrons
localize, and on the eventual influence of the atomic spin-orbit
coupling. Taking the example of $d^4$ ions, for large Hund's exchange
high-spin $S=2$ states emerge at Mn ions in LaMnO$_3$
\cite{Fei99,Dag04,Kov10,Sna18}, and the corresponding $A$-type
antiferromagnetic ($A$-AF) spin order follows the Goodenough-Kanamori
rules \cite{Goode}. On the other hand, when Hund's exchange is weaker
compared with $t_{2g}$--$e_g$ splitting, $e_g$ orbitals are empty and
intermediate $S=1$ spin states form at $d^4$ ions. An example of such
a $t_{2g}$ system is Ca$_2$RuO$_4$ with spin-orbital superexchange
described in section 2. Here, the orbital degree of freedom is set by
a doubly occupied orbital configuration called \textit{doublon}
\cite{Fiona,Cuo06,Brz15}, see Figs. 1(b-d), where we introduce the
following notation to identify which orbital of the $t_{2g}$ sector
is doubly occupied \cite{Kha05},
\begin{equation}
\left|a\right\rangle\equiv\left|yz\right\rangle, \quad
\left|b\right\rangle\equiv\left|xz\right\rangle, \quad
\left|c\right\rangle\equiv\left|xy\right\rangle.
\label{eq:orbs}
\end{equation}
Without distortions, only two out of three $t_{2g}$ orbitals are active
along each bond $\langle ij\rangle\parallel\gamma$ and contribute to
the intersite kinetic energy, while the third orbital is inactive as
the hopping via oxygen is forbidden by symmetry. The inactive $t_{2g}$
orbital along a given cubic axis $\gamma\in\{a,b,c\}$ is in the basis
(\ref{eq:orbs}) uniquely labeled as $|\gamma\rangle$.

\begin{figure}[t!]
\begin{center}
\includegraphics[width=\columnwidth]{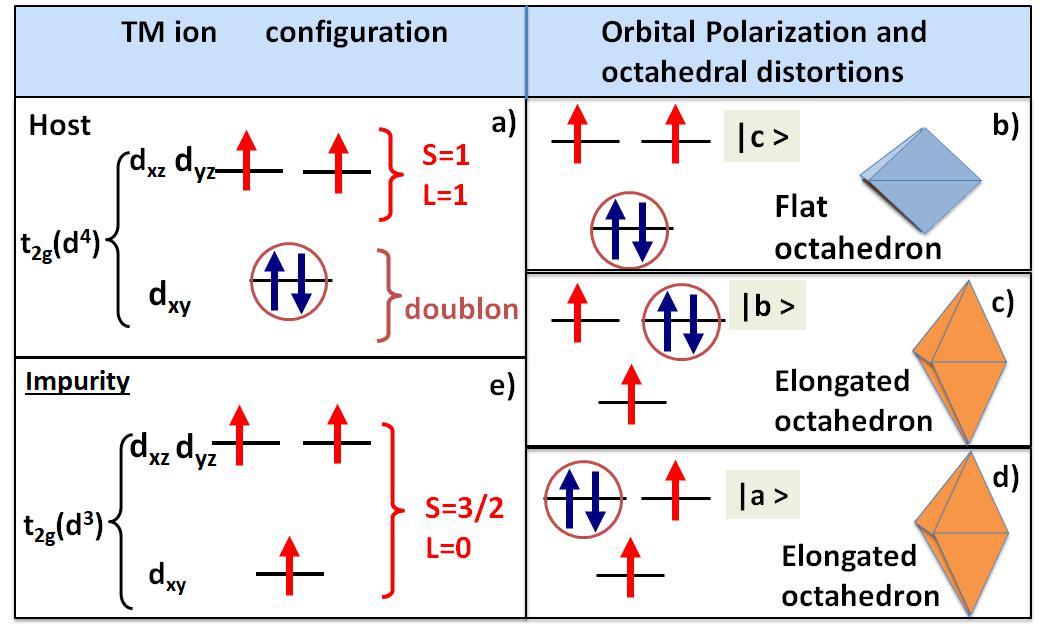}
\end{center}
\caption{Artist's view of the orbital doping when:
(a) $d^4$ Ru ion with spin $S=1$ and three $t_{2g}$ orbitals split
by the crystal field, with $L=1$ angular momentum,
(e) is substituted by the $d^3$ Cr ion without orbital degree of
freedom ($L=0$) and spin $S=3/2$ due to Hund's exchange.
Spins are shown by red arrows and doubly occupied $t_{2g}$ orbitals
(doublons) are indicated by blue arrows. Depending on the
octahedral distortion, the doublon occupies:
(b) $|c\rangle$ orbital for a flat octahedron, or
(c,d) one of two degenerate doublet $\{|a\rangle,|b\rangle\}$ orbitals
for an elongated octahedron.}
\label{fig:1}
\end{figure}

Doping Mott insulators typically refers to the addition of charge
carriers that leads to melting of the localized spin-orbital order
\cite{Lee06,Dag01} and the formation of metallic states often
accompanied by unconventional superconductivity, to novel patterns
(stripes, nematic, \textit{etcetera} \cite{Voj09}) of self-organized
electronic states \cite{Eme97,Wro10,Prl15}, or to the suppression of
orbital order by the orbital rotation induced by charge defects
\cite{Ave19}. In contrast, substitutional doping of transition metal
elements can avoid the breakdown of the Mott insulating state and
allows novel directions for tuning the degree of intertwining of the
spin-orbital correlations. As an experimental motivation we mention the
anomalous magnetic reconstruction that realizes $3d^3$ substitution in
a $4d^4$ host, or Mn doping in Sr$_3$Ru$_2$O$_7$ \cite{Mes12}.

The potential to modify the valence of the transition metal elements
without destroying the insulating state also allows to achieve
non-standard regimes of competition between lattice and spin-orbital
degrees of freedom. Indeed, specific for the present study is the
physical case of the Mott insulating Ca$_2$RuO$_4$ whose substitution
of Ru with Mn \cite{Qi12}, Cr \cite{Qi10}, Fe \cite{Qi12}, or Ir
\cite{Yuan15} atoms, apart from driving a modification of the
spin-orbital order, leads to the observation of negative thermal
expansion effects, i.e., the increase of the unit cell volume by
thermal cooling. These examples suggest that the negative thermal
expansion outcomes are strongly linked to the electronic correlations
developing in the doped Mott phase.

\begin{figure}[t!]
\begin{center}
\includegraphics[width=\columnwidth]{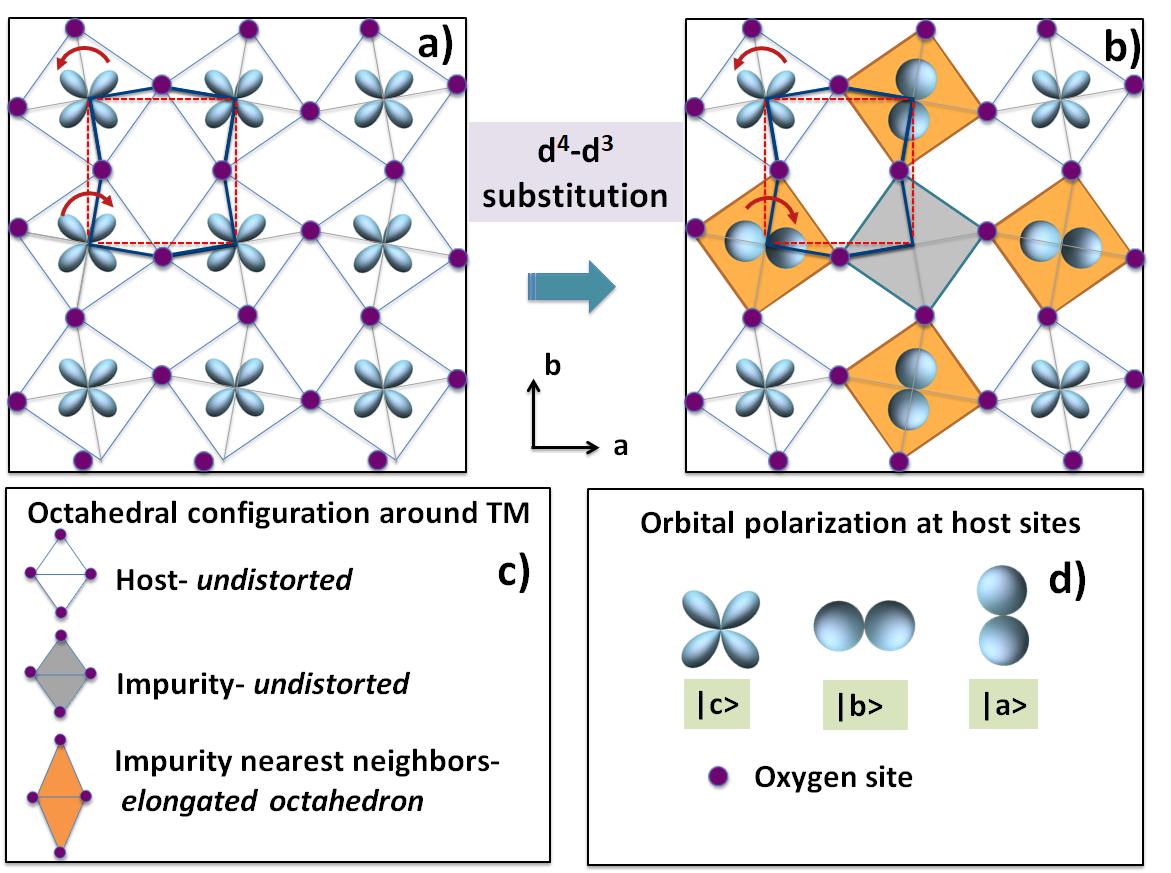}
\end{center}
\caption{Artist's view of the evolution of orbital order from
(a) the undoped $(a,b)$ plane of $d^4$ ions in centers of
rotated octahedra (arrows) forming a two-sublattice pattern,
(b) the plane with a $d^3$ ion in the center, and
(c) undistorted (elongated) octahedra at (around) the impurity,
respectively, shown by gray (orange) color.
The undoped state (a) has a uniform ferro-orbital (FO) order with the
doublon occupying the $|c\rangle$ orbital.
Panel (d) shows the possible orbital configurations (\ref{eq:orbs}) of
the doublon at an undoped $d^4$ site, see Fig. \ref{fig:1}. Finally,
the violet circles stand for oxygen sites in the $(a,b)$ plane.
}
\label{fig:2}
\end{figure}

The doping path, indeed, makes possible the design of spin-orbital
defects that in turn are expected to cause significant deviations from
the standard exchange as related to the Goodenough-Kanamori rules
\cite{Brz15,Brz16,Brz17,Brz18}. For instance, focusing on a TMO with
$t_{2g}$ orbital degrees of freedom, one can achieve orbital dilution
by doping with orbitally inactive transition metal magnetic ions
\cite{Brz15}. Such doping of $t_{2g}$ sector can be realized by
replacing a $d^4$ Ru ion with low $S=1$ spin and effective $L=1$
angular momentum with a $d^3$ ion corresponding with a local increase
of spin to $S=\frac{3}{2}$ and removal of the orbital degree of freedom
(i.e., $L=0$), see Figs. \ref{fig:1}(a) and \ref{fig:1}(e). Alternative
substitutional doping by $d^2$ ions employs both the orbital and charge
degrees of freedom and is called charge dilution \cite{Brz17,Brz18}.

Focusing on $d^3$ substitution in $d^4$ systems, the im\-purity-host
spin-orbital exchange is strongly dependent on the orbital that is
doubly occupied in the $t_{2g}$ sector. Indeed, the
impurity can act as an orbital vacancy for weak host-impurity coupling
or favor an orbital polaronic configuration with the doublon sitting
in the active orbitals along the $3d-4d$ bond \cite{Brz15}. For an
$(a,b)$ plane, the orbital selection of the doublon configuration
couples in a different way to the octahedral distortions corresponding
to flat or elongated modes, see Figs. \ref{fig:1}(b)--\ref{fig:1}(d).
Thus we explore the interrelation between the orbital order developing
around the impurity and the character of the compatible octahedral
distortions. We find that the rotation of the octahedra is able to pin
uniquely the type of orbital order around the impurity (see Fig. 2).
This result indicates a strong bias in the induced octahedral
distortions. The main outcomes of the present investigation include:
\hfill\break
($i$) the determination of the effective $d^3-d^4$ superexchange in the
presence of octahedral rotations, \hfill\break
($ii$) establishing orbital order around the impurity, \hfill\break
($iii$) providing a discussion on the way the orbital order is linked
to the flat or elongated distortions of the octahedra in the host.

\section{Spin-orbital superexchange in the host}
\label{sec:od}

First we consider the effective spin-orbital superexchange in the host
taking the limit of strong local Coulomb interactions $U_2$ at Ru ions
in $4d^4$ local configurations, i.e., charge excitations
$d^4_id^4_j\rightarrow d^5_id^3_j$, along each bond $\langle ij\rangle$
generate effective superexchange \cite{Fiona,Cuo06,Brz15},
\begin{equation}
{\cal H}_{d^4-d^4}=J_{\rm host}\sum_{\langle ij\rangle\parallel\gamma}
\left\{{\cal J}_{ij}^{(\gamma)}\left(\vec{S}_i\cdot\vec{S}_j+1\right)
+{\cal K}_{ij}^{(\gamma)}\right\}.
\label{eq:host}
\end{equation}
Here superexchange $\propto J_{\rm host}$ involves spin operators for
$S=1$ spins and includes on the orbital operators,
${\cal J}_{ij}^{(\gamma)}$ and ${\cal K}_{ij}^{(\gamma)}$,
which depend on the orbital doublet active along the bond direction
$\gamma$ ($\langle ij\rangle\parallel\gamma$). The superexchange
(\ref{eq:host}) may be obtained from that for vanadium perovskites
with V $d^2$ ions \cite{Kha01,Ole07} by electron-hole transformation,
where an empty site (holon) transforms into a doublon for $d^4$ ions.
The form of $\{{\cal J}_{ij}^{(\gamma)},{\cal K}_{ij}^{(\gamma)}\}$ was
given in \mbox{\cite{Kha01,Ole07,Hor08}}; $J_{\rm host}$ depends on the
hopping $t$ and on Coulomb $U_2$ and Hund's exchange $J_2^H$ host
parameters, and $\eta_{\rm host}\equiv J_2^H/U_2$. An additional
aspect which we neglect here is that spin-orbital entangled variables
that drive magnetism in ruthenates are influenced by electron-lattice
coupling, for instance via pseudo-Jahn-Teller effect \cite{Liu19},
which focuses on the distortions which split the $t_{2g}$ orbitals.

We consider here a 2D square lattice with transition metal ions
connected via oxygen ions as in an $(a,b)$ RuO$_2$ plane, see Fig.
\ref{fig:2}, of Ca$_2$RuO$_4$. In this case $|a\rangle$ ($|b\rangle$)
orbitals are active along the $b$ ($a$) axis, while $|c\rangle$
orbitals are active along both $a,b$ axes. Finite crystal field (CF)
favors doublons in $|c\rangle$ orbitals, see Fig. 2(a). Below we
investigate the effect of orbital dilution shown in Fig. 2(b).

For the host sites we assume AF spin order replacing spin-spin
interactions by the correlation function
$\langle{\bf S}_{i}{\bf S}_{j}\rangle\simeq-5/4$
(here we neglect spin quantum fluctuations)
and add the anisotropic spin-orbit coupling term in a form of,
\[
{\cal H}_{so}=\lambda\sum_{i}(-1)^{i}h_{z}L_{i}^{z},
\]
where $h_{z}$ is the local staggered magnetic moment assumed to be
along $z$ axis and the orbital $L_i^z$ operator has a standard form
of $L_{i}^{z}=(ia_{i}^{\dagger}b_{i}+H.c.)$. Altogether, the total
host Hamiltonian reads,
${\cal H}_{\rm host}={\cal H}_{d^4-d^4}+{\cal H}_{so}$.

\section{Hybrid bond: Orbital dilution}
\label{sec:hyb}

In this section we present the results of the derivation of effective
$3d^3-4d^4$ spin-orbital superexchange as due to the coupling between
orbitals of $3d$ and $4d$ ions through the oxygen $2p$ orbitals which
build up the $p-d$ hybridization $\propto  V_{pd\pi}^2$. For our
purposes, it is sufficient to analyze a pair of atoms forming a bond
$\langle ij\rangle$, as the effective interactions are generated by
charge excitations, $d^n_id^m_j\rightarrow d^{(n+1)}_i d^{(m-1)}_j$
along a single bond \cite{Ole05}. In contrast to the reference $d^4$
host where both atoms on the bond $\langle ij\rangle$ are equivalent,
a $d^3-d^4$ hybrid bond has explicitly different ionic configurations.
The degenerate Hubbard Hamiltonian $H(i,j)$ \cite{Ole83} includes in
general the standard local Coulomb interaction $H_{\rm int}(i)$ and the
effective $d-d$ kinetic term, $H_t(i,j)$; for a representative
$3d$-$2p$-$4d$ bond after projecting out the oxygen degrees of freedom,
\begin{equation}
H(i,j)=H_t(i,j)+H_{\rm int}(i)+H_{\rm int}(j) \,.
\label{host}
\end{equation}
The case without octahedral rotation was investigated in \cite{Brz15}.
Here we present the superexchange Hamiltonian for a bond
$\langle 12\rangle$ along the $a$ axis between the $d^3$ impurity at
site $i=1$ and a host $d^4$ ion at site $j=2$, in presence of
octahedral rotation by angle $\phi$,
\begin{eqnarray}
&&{\cal H}_{d^3-d^4}^{(a)}(\phi)=
J_{\rm imp}\left({\bf S}_1\cdot{\bf S}_2\right) \nonumber\\
&\times&\left\{\alpha_1+\alpha_2\left(a_2^{\dagger}b_2^{}+b_2^{\dagger}a_2^{}\right)
+\alpha_3a_2^{\dagger}a_2+\alpha_4b_2^{\dagger}b_2^{}\right\}  \nonumber\\ &+&\left\{\beta_{1}+\beta_2\left(a_2^{\dagger}b_2^{}+b_2^{\dagger}a_2^{}\right)
+\beta_3a_2^{\dagger}a_2^{}+\beta_4b_2^{\dagger}b_2^{}\right\}.
\label{eq:d3-d4}
\end{eqnarray}
Here the coefficients $\{\alpha_i\}$ and $\{\beta_i\}$ are given by
the Slater-Koster rules \cite{Slako}, with the property that
$\{\alpha_2,\alpha_4\}$ and $\{\beta_2,\beta_4\}$ vanish for $\phi=0$,
so they are generated by the rotation. At $\phi=0$, the Hamiltonian
tends to project out the inactive orbital $a$ along the bond
$\langle ij\rangle\parallel a$ axis. With angle $\phi\neq 0$, also
the $c$ orbital is disfavored so that only the $b$ orbital is the
preferred one. In this way we obtain polarization of the orbitals
$\{a,b\}$ towards the impurity --- the orbital polarizer mechanism,
see Fig. 2.

The exact form of the coefficients $\{\alpha_i\}$ and $\{\beta_i\}$ as
function of rotation angle $\phi$ reads as
\begin{eqnarray}
\alpha_1&=&-\frac{2}{9}\gamma+\frac{1}{6}q_{5}+\frac{4\gamma+3}{18}q_3,
\nonumber\\
\alpha_2&=&-\sin(2\phi)\left(-\frac{1}{9}-\frac{1}{12}q_5+\frac{1}{36}q_3\right),
\nonumber\\
\alpha_3&=&\left(\sin^{2}\phi-\gamma\right)\left(
-\frac{2}{9}-\frac{1}{6}q_{5}+\frac{1}{18}q_{3}\right),       \nonumber\\
\alpha_4&=&\left(\cos^{2}\phi-\gamma\right)\left(
-\frac{2}{9}-\frac{1}{6}q_{5}+\frac{1}{18}q_{3}\right),
\end{eqnarray}
and
\begin{eqnarray}
\beta_1&=&-\frac23\gamma-\frac14 q_5-\frac{4\gamma+3}{12}q_3, \nonumber\\
\beta_2&=&-\sin(2\phi)\left(-\frac13 +\frac18 q_5-\frac{1}{24}q_3\right),\nonumber\\
\beta_3&=&\left(\sin^{2}\phi-\gamma\right)\left(
-\frac{2}{3}+\frac{1}{4}q_{5}-\frac{1}{12}q_{3}\right), \nonumber\\
\beta_4&=&\left(\cos^{2}\phi-\gamma\right)\left(
-\frac{2}{3}+\frac{1}{4}q_{5}-\frac{1}{12}q_{3}\right),
\end{eqnarray}
with dimensionless parameters,
\begin{equation}
q_{i}=\frac{1}{i\eta_{\rm imp}+1}\,,\quad
\gamma=\left(\frac{t_{c,c}^{(a)}}{{\tilde V}_{pd\pi}^2}\right)^2,\quad
\eta_{\rm imp}=\frac{J_{1}^{H}}{\Delta}.
\end{equation}

The $d^3-d^4$ superexchange (\ref{eq:d3-d4}), is given by,
\begin{equation}
J_{\rm imp}\!=\frac{4\tilde{V}_{pd\pi}^{4}}{\Delta}, \hskip .2cm
\Delta=I_{e}+3(U_{1}+U_{2})-4(J_{1}^{H}-J_{2}^{H}),
\end{equation}
and is determined by the effective hopping between two neighboring
sites in a two-step process $\propto V_{pd\pi}^2$, involving a
charge-transfer excitation energy $\Delta_{\rm CT}$ along the
$d$-$p$-$d$ $\pi$-bond. We introduce for convenience the energy,
\mbox{$\tilde{V}_{pd\pi}^2\equiv V_{pd\pi}^2/\Delta_{\rm CT}$};
Coulomb and Hund's exchange parameters are: $\{U_1,J_1^H\}$. The
hopping amplitudes are given by the Slater-Koster rules \cite{Slako},
\begin{eqnarray}
t_{c,c}^{(a)}&=&-{\tilde V}_{pd\pi}^{2}\cos^{3}(2\phi)  \nonumber\\
&+&\frac{1}{8}\Big\{{\tilde V}_{pd\sigma}\!\left(3{\tilde V}_{pd\sigma}
-4\sqrt{3}{\tilde V}_{pd\pi}\right)\sin(2\phi)\sin(4\phi)\Big\}, \\
t_{a,a}^{(a)} & = & \quad\! \tilde{V}_{pd\pi}^{2}\sin^{2}\phi,  \\
t_{b,b}^{(a)} & = & -\tilde{V}_{pd\pi}^{2}\cos^{2}\phi,         \\
t_{a,b}^{(a)} & = & -\tilde{V}_{pd\pi}^{2}\sin\phi\cos\phi.
\end{eqnarray}
Here we take a realistic assumption,
${\tilde V}_{pd\sigma}=2{\tilde V}_{pd\pi}$. To obtain the superexchange
(\ref{eq:d3-d4}) for a $b$ bond, we transform the coefficients as follows,
\begin{equation}
\alpha_{1}\to \,\alpha_{1},  \quad
\alpha_{2}\to  -\alpha_{2},  \quad
\alpha_{3}\to \,\alpha_{4},  \quad
\alpha_{4}\to \,\alpha_{3},
\end{equation}
and analogous for the $\{\beta_i\}$ ones.

\section{Results: Crystal field modification}
\label{sec:res}

\begin{figure}[t!]
\begin{center}
\includegraphics[width=.77\columnwidth]{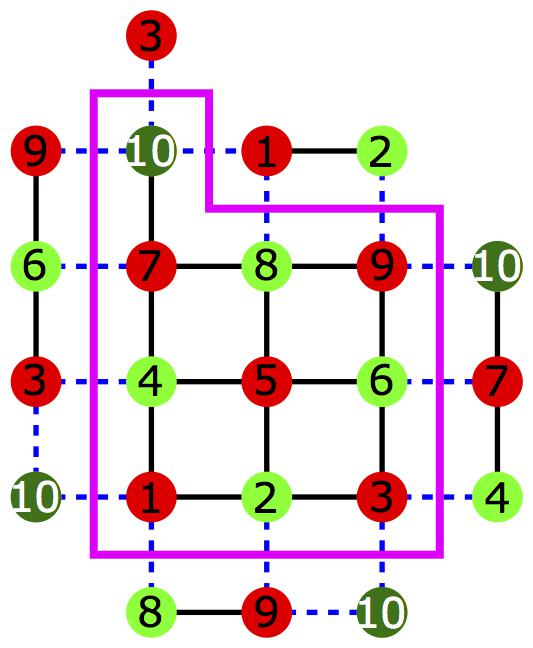}
\end{center}
\caption{
Schematic view of a 10-site periodic cluster (magenta boarder line)
used for calculation of the crystal field at the nearest neighbors of
$d^3$ impurity ($i\in\Omega$, $i=1,3,7,9$), with impurity at $i=10$
and $S=\frac32$. Orbital degrees of freedom remain undisturbed at next
nearest neighbors ($i=2,4,6,8$) and the only third nearest neighbor
$i=5$ in the cluster.
Two-sublattice AF N\'eel spin order is indicated by green and red sites.
}
\label{fig:3}
\end{figure}

For the exact diagonalization we use a periodic cluster of 10 sites
with one impurity at site $i=10$, see Fig. \ref{fig:3}. Since we want to
focus on the effect of the impurity, we take it in the strong coupling
regime, $J_{\rm imp}\gg J_{\rm host}$. We neglect the quantum spin
fluctuations and thus the size of the Hilbert space becomes
computationally accessible. We consider a collinear AF state
described by Ising variables, i.e., we assume the N\'eel AF spin order.
The orbital order in the host is ferro-orbital (FO), with $|c\rangle$
doublons, see Fig. \ref{fig:2}(a). The impurity disturbs the orbitals
at its neighbors and generates the effective change of the CF due to
correlations at the host sites $i\in\Omega$. As a result, for
small $\Delta_i^c$ the doublons occupy directional orbitals pointing
towards the impurity, similar to the orbital polarons in manganites
\cite{Kha99,Ake03,Gec05}.

\begin{figure}[t!]
\begin{center}
\includegraphics[width=.97\columnwidth]{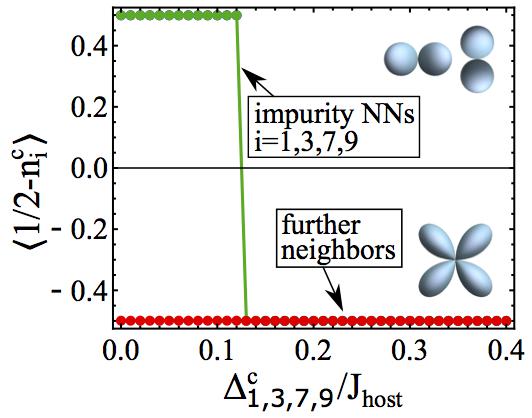}
\end{center}
\caption{
Occupation of the $c$ orbital at site $i$ of the cluster shown in Fig.
\ref{fig:3} as a measure of orbital order for $\phi=0.25$. The average
doublon occupancy $\langle n_i^c\rangle$ changes at site $i$ being a
nearest neighbor (NN) of the impurity. For increasing value of the CF
splitting at NNs of impurity $\Delta^c$, the doublon moves from a
directional orbital, $a$ or $b$, pointing towards the impurity to
an in-plane $|c\rangle$ orbital (see insets) and the FO order is
restored ($\langle\frac12-n_i^c\rangle=\mp\frac12$ stands for site
$i$ unoccupied/occupied by the $c$ doublon).
Energy parameters (all in units of $J_{\rm host}$)):
$\lambda=0.1$, $J_{\rm imp}=8.0$; other parameters:
$\eta_{\rm host}=0.05$, $\eta_{\rm imp}=0.2$,
$\langle\vec{S}_i\cdot\vec{S}_{10}\rangle=-2.0$ with $i\in\Omega$.
}
\label{fig:4}
\end{figure}

To estimate the strength of this inverted CF, we put a finite CF
$\Delta_i^c$ at nearest neighbor sites of the impurity and find
that for large enough $\Delta_i^c$ the doublon orbital
changes from $|a\rangle$ or $|b\rangle$ to $|c\rangle$, i.e.,
to the one along the bond, see \ref{fig:2}(b).
The change of orbital occupation at sites around the impurity at
$\Delta_i^c\simeq 0.12J_{\rm host}$ is shown in Fig. \ref{fig:4}.
Remarkably, the bonds along the $a$ and $b$ axes are equivalent and we
observe a change of doublons from directional
($\Delta_i^c\le 0.12J_{\rm host}$) to planar
($\Delta_i^c> 0.12J_{\rm host}$) orbitals. The inversion of CF
occurs at the same value of $\Delta_i^c$ for the
bonds along the $a$ axis ($i=1,9$) and along the $b$ axis ($i=3,7$),
see Fig.~\ref{fig:4}. 

\begin{figure}[t!]
\begin{center}
\includegraphics[width=\columnwidth]{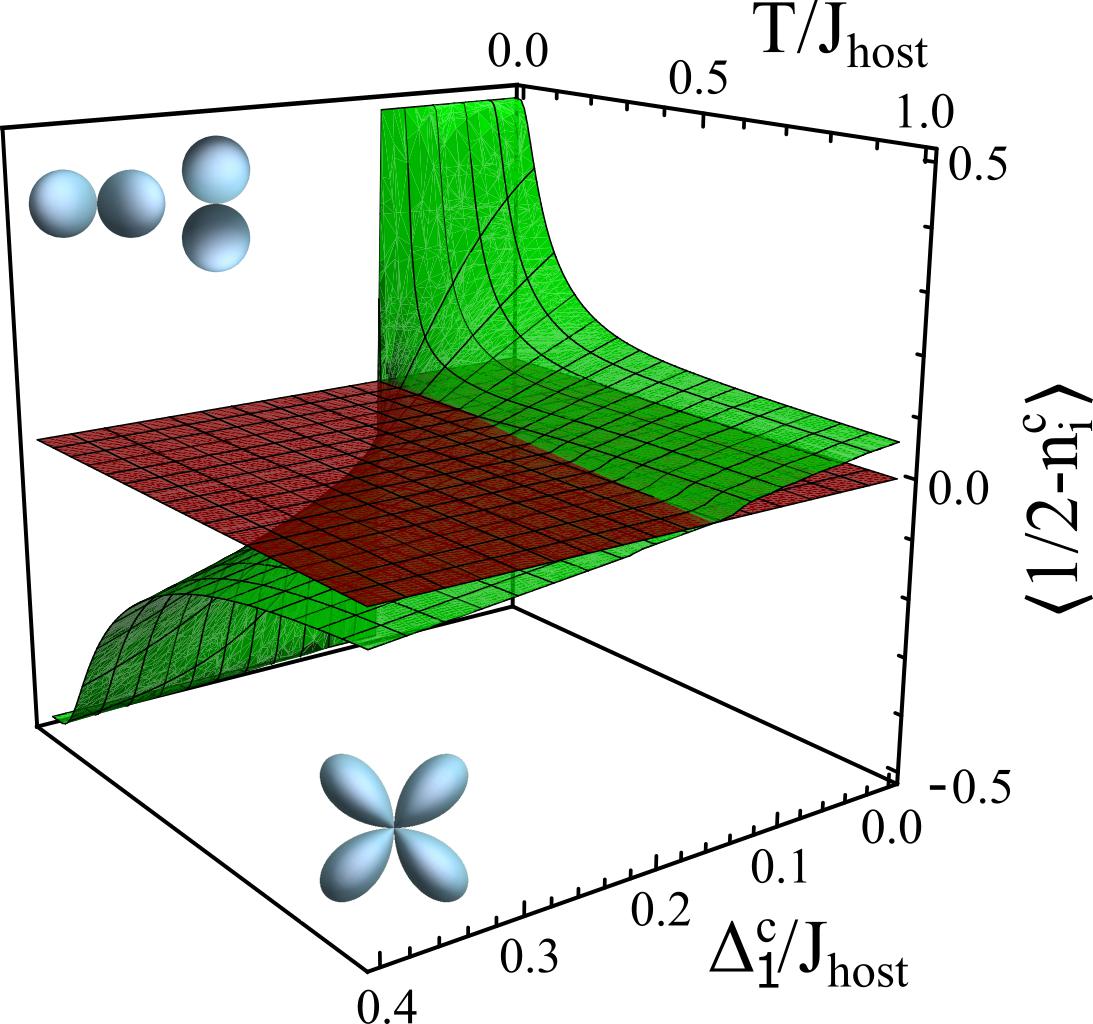}
\end{center}
\caption{
Evolution of the $c$ orbital occupation at the host site $i=1$ (green),
neighboring with impurity (see Fig. \ref{fig:3}), as function of CF
$\Delta_1^c$ and temperature $T$ [in units of $J_{\rm host}\equiv 1$].
The average doublon occupancy $\langle n_1^c\rangle$ changes with
increasing $\Delta_1^c$ or $T$. The plane (dark red) sets the zero
value for the $c$ orbital polarization and separates the regions of
$a/b$ doublon ($\langle n_1^c\rangle=0$) and $c$ doublon
($\langle n_1^c\rangle>0$).
Parameters as in Fig. \ref{fig:4}.
}
\label{fig:5}
\end{figure}

An abrupt change of the doublon orbital is modified to a smooth
crossover at finite temperature $T$, see Fig. \ref{fig:5}. Thermal
fluctuations generate a rather broad range of CF splitting
$\Delta_1^c$, where $\langle\frac12-n_1^c\rangle\simeq 0$
and this quantity changes sign close to the value
$\Delta_1^c\simeq 0.14J_{\rm host}$ found before at $T=0$, see Fig.
\ref{fig:4}. Interestingly, we find that the strength of the effective
CF potential induced by exchange interaction is weakly dependent on
temperature, so that such electronically driven orbital
splitting is remarkably robust against thermal fluctuations.

\section{Discussion and Conclusions}
\label{sec:summa}

We have shown that orbital doping in the presence of octahedral
rotations around the $c$ axis tends to favor a distinct type of orbital
order, with orbital polarization that is preferentially directional and
distributed both towards the impurity and out-of-plane (i.e., with
either $xz$ or $yz$ orbital symmetry). Remarkably, as already
demonstrated in the tetragonal symmetric octahedra \cite{Brz15}, the
pinning of the orbital order can occur for both an antiferromagnetic
and ferromagnetic exchange between host and dopants. Its manifestation
depends either on the amplitude of Hund's exchange coupling or on the
relative strength of the host-host to host-impurity exchange
interactions. This means that the local orbital order is a generic sign
of the $d^3$ dopant in a distorted host with octahedral rotations and
is robust to spin fluctuations.

Another relevant and striking consequence of the specific orbital order
induced locally by the $d^3$ dopant is that the orbital pattern around
the impurity is uniquely compatible with an elongated octahedral
configuration, see Fig. 2. Hence, the impurity-host exchange yields an
effective crystal field potential that is akin to that obtained when
the lattice favors longer out-of-plane transition metal--oxygen bonds
than the in--plane ones. On this basis, there are two possible emergent
physical scenarios that can occur: \hfill\break
($i$) If the lattice potential stabilizes an octahedral
configuration that is flat, then, the host-impurity exchange tends to
compete with it and, depending on their relative strength, can even end
up reversing the sign of the crystal field interaction. This is the
case we demonstrate in this paper. Such occurrence clearly implies that
the optimal local deformation exhibits an
effective enhanced volume of the unit cell due to the bond expansion of
the octahedra along the $c$ axis that, in turn, can play a relevant role
in setting non-standard negative thermal expansion effects once orbital
order is achieved at low temperature. \hfill\break
($ii$) On the contrary, for a host configuration with elongated
octahedra, the effective host-impurity exchange can enhance the
distortions around the impurity thus increasing the stiffness of the
lattice. In the present paper, the analysis has been motivated by the
study of $d^3$ dopants in $d^4$ Mott insulating host with flat
octahedra and antiferromagnetic order as it occurs in Mn-doped
Ca$_2$RuO$_4$ compound \cite{Qi12}. Hence, we speculate that the
outcome of the induced elongated octahedra by Coulomb driven orbital
exchange may be relevant for the anomalous volume expansion
occurring below the onset temperatures of magnetic and orbital
order~\cite{Qi12}.

It is interesting to point out that similar competing effects between
electron correlations and lattice distortions do not occur in transition
metal oxides with $e_g$ orbital degrees of freedom, whereas the
Jahn-Teller distortions are typically cooperating with the orbital
exchange of electronic origin \cite{Fei99,Kha99,Oka02,Cuo02}. The local
lattice distortions associated with the variation of the orbital order
at the metal-insulator transition have been detected by EXAFS and
XANES in manganites \cite{Lan98} and recently in PbTiO$_3$-based
perovskites systems \cite{Pan19}. Therefore, we expect that further
experiments will provide interesting information in this field.

We also underline that the temperature dependence of the Coulomb driven
effective crystal field (Fig. 5) allows one to have an anomalous thermal
behavior when considering the thermal expansion effects. Indeed, since
the obtained results show that the amplitude of the effective exchange
driven crystal field potential is reduced by the increase of the
temperature, then, one can achieve a regime of anomalous volume
expansion only in a given window of temperature. This is, for instance,
the case of Cr-doped Ca$_2$RuO$_4$ \cite{Qi10} where the negative
thermal effects manifest predominantly in a range of temperature below
the magnetic transition.

Finally, we point out that the proposed reconstruction of the orbital
order around the dopant can be accessed by experimental probes based on
X-ray spectroscopy which is element sensitive and has been successfully
demonstrated to unveil the character of the ordered ground state and
the corresponding spin-orbital excitations in such transition metal
oxides \cite{Das2018,Porter2018}.

\noindent\textbf{Acknowledgments}
Open access funding provided by Max Planck Society.
We acknowledge support by Naro\-dowe Centrum Nauki (NCN,
National Science Centre, Poland), Project No. 2016/23/B/ST3/00839.\\
A.~M.~Ole\'s is grateful for the Alexander von Humboldt Foundation
Fellowship \mbox{(Humboldt-Forschungspreis).}

\end{document}